\documentclass[11pt]{article}
\pdfoutput=1
\usepackage{fullpage}
\usepackage{xcolor}
\usepackage{cite}
\usepackage{graphicx}
\usepackage{amsmath}
\usepackage{amssymb}
\usepackage{subcaption}
\usepackage{stackrel}
\usepackage[utf8x]{inputenc} 
\title{}
\date{}

\def\beq{\begin{equation}}
\def\eeq{\end{equation}}
\allowdisplaybreaks
\begin{document}
\bibliographystyle{utphys}

\newcommand{\be}{\begin{equation}}
\newcommand{\ee}{\end{equation}}
\newcommand\n[1]{\textcolor{red}{(#1)}} 
\newcommand{\diff}{\mathop{}\!\mathrm{d}}
\newcommand{\lb}{\left}
\newcommand{\rb}{\right}
\newcommand{\f}{\frac}
\newcommand{\pd}{\partial}
\newcommand{\tr}{\text{tr}}
\newcommand{\fdiff}{\mathcal{D}}
\newcommand{\im}{\text{im}}
\let\caron\v
\renewcommand{\v}{\mathbf}
\newcommand{\T}{\tensor}
\newcommand{\R}{\mathbb{R}}
\newcommand{\C}{\mathbb{C}}
\newcommand{\Z}{\mathbb{Z}}
\newcommand{\msbar}{\ensuremath{\overline{\text{MS}}}}
\newcommand{\DIS}{\ensuremath{\text{DIS}}}
\newcommand{\abar}{\ensuremath{\bar{\alpha}_S}}
\newcommand{\bb}{\ensuremath{\bar{\beta}_0}}
\newcommand{\rc}{\ensuremath{r_{\text{cut}}}}
\newcommand{\Nd}{\ensuremath{N_{\text{d.o.f.}}}}
\newcommand{\red}[1]{{\color{red} #1}}
\setlength{\parindent}{0pt}

\titlepage
\begin{flushright}
QMUL-PH-21-35\\
\end{flushright}

\vspace*{0.5cm}

\begin{center}
{\bf \Large The double copy of the multipole expansion}

\vspace*{1cm} 
\textsc{Erick Chac\'{o}n$^a$\footnote{e.c.chaconramirez@qmul.ac.uk},
Andr\'{e}s Luna$^b$\footnote{luna@physics.ucla.edu},
  and Chris D. White$^a$\footnote{christopher.white@qmul.ac.uk}} \\

\vspace*{0.5cm} Centre for Research in String Theory, School of
Physics and Astronomy, \\
Queen Mary University of London, 327 Mile End
Road, London E1 4NS, UK\\

\vspace*{0.5cm} Mani L. Bhaumik Institute for Theoretical Physics,
Department of Physics and Astronomy, UCLA, Los Angeles, CA 90095\\

\end{center}

\vspace*{0.5cm}

\begin{abstract}
We consider the classical double copy, that relates solutions of
biadjoint scalar, gauge and gravity theories. Using a recently
developed twistor expression of this idea, we use well-established
techniques to show that the multipole moments of arbitrary vacuum type
D gravity fields are straightforwardly mapped to their counterparts in
gauge and biadjoint scalar theory by the single and zeroth copies. We
cross-check our results using previously obtained results for the Kerr
metric. Our results provide further physical intuition of how the
double copy operates.
\end{abstract}

\vspace*{0.5cm}

\section{Introduction}
\label{sec:intro}

There is mounting evidence that our various theories of fundamental
physics are more closely connected than previously thought. In this
paper, we will focus on a particular correspondence -- the {\it
  classical double copy} -- that relates solutions of the field
equations in (non-)abelian gauge theories and gravity, as well as in a
novel scalar theory with two different types of colour charge ({\it
  biadjoint scalar field theory}). Inspired by the original double
copy for scattering amplitudes in the corresponding quantum field
theories~\cite{Bern:2010ue,Bern:2010yg} (which itself has a string
theoretic origin~\cite{Kawai:1985xq}), the first classical double copy
to appear was the {\it Kerr-Schild double copy} of
ref.~\cite{Monteiro:2014cda} (see
refs.~\cite{Luna:2015paa,Ridgway:2015fdl,Bahjat-Abbas:2017htu,Berman:2018hwd,Carrillo-Gonzalez:2017iyj,CarrilloGonzalez:2019gof,Bah:2019sda,Lescano:2020nve,Gumus:2020hbb,Pasterski:2020pdk,Berman:2020xvs,Easson:2020esh,Alkac:2021bav,Alkac:2021seh}
for further developments). An alternative exact double copy procedure is
the {\it Weyl double copy} of ref.~\cite{Luna:2018dpt} (see also
refs.~\cite{Keeler:2020rcv,Pasterski:2020pdk,Sabharwal:2019ngs,Alawadhi:2020jrv,Godazgar:2020zbv}). This
uses the spinorial rather than tensorial formalism of General
Relativity, and includes the Kerr-Schild double copy as a special
case. To date, it constitutes the most general exact statement of the
classical double copy, although other formalisms also offer useful
alternative
insights~\cite{Elor:2020nqe,Farnsworth:2021wvs,Anastasiou:2014qba,LopesCardoso:2018xes,Anastasiou:2018rdx,Chacon:2020fmr,Luna:2020adi,Borsten:2020xbt,Borsten:2020zgj,Borsten:2021hua,Cheung:2020djz,Cheung:2021zvb}.\\

As well as practical applications of this correspondence (see
e.g. ref.~\cite{Bern:2019prr}), there are also important conceptual
issues to address, including understanding the ultimate origin of the
classical double copy itself. To this end,
refs.~\cite{White:2020sfn,Chacon:2021wbr} recently showed how one can
derive both the form and scope of the Weyl double copy using
well-established ideas from twistor
theory~\cite{Penrose:1967wn,Penrose:1972ia,Penrose:1968me} (see
refs.~\cite{Penrose:1987uia,Penrose:1986ca,Huggett:1986fs} for
pedagogical reviews of this subject), as well as showing that the Weyl
copy is more general than previously thought. As well as applying new
mathematical techniques, another useful method for extending our
understanding of the double copy is to take known physical or
mathematical properties in biadjoint scalar, gauge and gravity
theories, and to see how they match up (or otherwise). Recent examples
include properties of solutions at strong
coupling~\cite{White:2016jzc,DeSmet:2017rve,Bahjat-Abbas:2018vgo,Bahjat-Abbas:2020cyb},
symmetries~\cite{Monteiro:2011pc,Borsten:2021hua,Alawadhi:2019urr,Banerjee:2019saj,Huang:2019cja},
and geometric / topological
information~\cite{Berman:2018hwd,Alfonsi:2020lub,Alawadhi:2021uie}. Even
more simply, the original Kerr-Schild double copy of
ref.~\cite{Monteiro:2014cda} told us that mass and energy map to
charge in the gauge theory, which nicely mirrors the replacement of
kinematic by colour information in scattering
amplitudes~\cite{Bern:2010ue,Bern:2010yg}.\\

In this paper, we extend these ideas by considering the well-known
{\it multipole expansion} of classical solutions. For any given
solution, one may define a series of higher-rank {\it multipole
  tensors}, which completely characterise the spatial and temporal
distribution of charge (energy / momentum) in the gauge (gravity)
theory respectively. The lowest-order results in this expansion
constitute the total charge and mass mentioned above, and one may then
ask whether higher-multipole moments are strict double copies of each
other. It turns out that this can be easily addressed in the twistor
picture of refs.~\cite{White:2020sfn,Chacon:2021wbr}. As has been
argued by Curtis~\cite{Curtis}, the multipole tensors can be replaced
by multi-index spinor fields, each of which satisfies the twistor
equation. As a consequence, one may instead describe the multipole
moments of a given field purely in terms of higher-rank twistors,
which are straightforwardly defined from the twistor-space functions
describing the fields. We will combine this with the twistor-space
double copy of refs.~\cite{White:2020sfn,Chacon:2021wbr}, and thus
obtain an explicit statement that the multipole expansion
double-copies, for arbitrary type $D$ vacuum solutions.  Our results
provide a useful physical insight into how the double copy operates,
and may well be relevant for thinking about further applications.\\


\section{The twistor space double copy}
\label{sec:twistor}

In this section, we review salient details of the Weyl double copy,
together with its twistor space incarnation, referring the reader to
refs.~\cite{White:2020sfn,Chacon:2021wbr} for full details. First, we
recall that massless free spacetime fields can be represented by
multi-index spinors $\phi_{AB\ldots C}$ ($\bar{\phi}_{A'B'\ldots
  C'}$), representing the anti-self-dual (self-dual) parts of the
field respectively. Index values run from 0 to 1, and may be raised,
lowered and / or contracted using the Levi-Civita symbols
$\epsilon^{AB}$ etc. There are $2n$ indices for a spin-$n$ field, and
the resulting quantities then satisfy a special case of the general
massless free field equation
\begin{equation}
\nabla^{AA'}\bar{\phi}_{A'\ldots C'}=0,\quad
\nabla^{AA'}\phi_{AB\ldots C}=0,
\label{masslessfreefield}
\end{equation}
where $\nabla^{AA'}$ is the appropriate translation of the spacetime
covariant derivative. These fields can be reinterpreted in twistor
space $\mathbb{T}$, corresponding to solutions of the {\it twistor
  equation}
\begin{equation}
\nabla_{A'}^{(A}\Omega^{B)}=0\quad\Rightarrow\quad \Omega^A=\omega^A
-ix^{AA'}\pi_{A'}.
\label{twistoreq}
\end{equation}
In the second equality we have written the general solution in
Minkowski space, in terms of constant spinors which may be grouped
together to make a 4-component {\it twistor}
\begin{equation}
Z^\alpha=(\omega^A,\pi_{A'}). 
\label{Zalpha}
\end{equation}
A non-local map between spacetime and twistor space is established by
requiring that the field in eq.~(\ref{twistoreq}) vanish, such that
the twistor components satisfy the incidence relation
\begin{equation}
  \omega^A=ix^{AA'}\pi_{A'}.
  \label{incidence}
\end{equation}
This is invariant under rescalings $Z^\alpha\rightarrow\lambda
Z^\alpha$, $\lambda\in{\mathbb C}$, such that we need only consider
{\it projective twistor space} $\mathbb{PT}$. A point in spacetime
corresponds to a Riemann sphere in $\mathbb{PT}$, also referred to as
a {\it (complex) line}. An important result known as the {\it Penrose
  transform} expresses massless free spacetime fields satisfying
eq.~(\ref{masslessfreefield}) via the contour integrals
\begin{equation}
  \bar{\phi}_{A'B'\ldots C'}(x)=\frac{1}{2\pi i}
  \oint_\Gamma d\pi_{E'}d\pi^{E'}\pi_{A'}\pi_{B'}\ldots \pi_{C'}
        [\rho_x f(Z^\alpha)],
  \label{Penrose}
\end{equation}
where $\rho_x$ restricts all twistors to obey the incidence relation
corresponding to spacetime point $x$, and the contour $\Gamma$ lies on
the appropriate Riemann sphere. The combined integrand and measure
must be invariant under rescalings $Z^\alpha\rightarrow \lambda
Z^\alpha$, which fixes the (holomorphic) function $f(Z^\alpha)$ to
have homogeneity $-(n+2)$ for a spin-$n$ field. The above remarks
imply that twistor functions of homogeneity $-2$, $-4$ and $-6$
correspond to spacetime gravity fields in scalar, gauge and gravity
theory respectively. Denoting the respective twistor functions by the
subscripts $\{{\rm scal.},{\rm EM},{\rm grav.}\}$ respectively,
refs.~\cite{White:2020sfn,Chacon:2021wbr} argued that one may define a
gravity twistor function via\footnote{We have here skimmed over the
fact that the twistor functions used throughout are not unique, and
are instead representatives of cohomology classes. The product of
eq.~(\ref{twistorcopy}) is then interpreted to apply only to
particular chosen representatives, as discussed in detail in
ref.~\cite{Chacon:2021wbr}.}
\begin{equation}
  f_{\rm grav.}(Z^\alpha)=\frac{f^{(1)}_{\rm EM}(Z^\alpha)
    f^{{(2)}}_{\rm EM}(Z^\alpha)}
  {f_{\rm scal.}(Z^\alpha)}, 
  \label{twistorcopy}
\end{equation}
leading to the spacetime {\it Weyl double copy}
formula
\begin{equation}
  \phi_{A'B'C'D'}(x)=\frac{\phi^{(1)}_{(A'B'}(x)\phi^{(2)}_{C'D')}(x)}
  {\phi(x)}
  \label{WeylDC}
\end{equation}
first presented in ref.~\cite{Luna:2018dpt}. Here $\phi$ is a
biadjoint scalar field, $\phi^{(i)}_{A'B'}$ an electromagnetic spinor,
and $\phi_{A'B'C'D}$ a Weyl spinor. The above discussion applies to
the case of primed spinor fields in spacetime. For unprimed fields,
one may consider the conjugate of eq.~(\ref{twistoreq}), whose
solutions are associated with {\it dual twistors} $W_\alpha$. The
notion of the Penrose transform can be straightforwardly adapted from
eq.~(\ref{Penrose2}):
\begin{equation}
  \phi_{AB\ldots C}(x)=\frac{1}{2\pi i}
  \oint_\Gamma d\lambda_{E}d\lambda^{E}\lambda_{A}\lambda_{B}\ldots
  \lambda_{C}
        [\rho_x f(W_\alpha)],
  \label{Penrose2}
\end{equation}
and the twistor double copy of eq.~(\ref{twistorcopy}) similarly
generalises. We will work with dual twistors by default in what
follows, in order to match conventions with
ref.~\cite{Curtis}.

  
\section{Multipoles and the double copy}
\label{sec:multipoles}

The idea of multipoles is familiar from Newtonian physics in
three-dimensional Euclidean space. A stationary electrostatic or
Newtonian potential $\phi$ in a sourceless region satisfies Laplace's
equation $\nabla^2\phi=0$, and may be expanded as\footnote{Throughout
  the paper we use lower-case Latin, upper-case Latin and Greek
  indices for spacetime tensors, spacetime spinors and twistors
  respectively. Note, however, that the indices in
  eq.~(\ref{newtonian}) run only over spatial components i.e. from 1
  to 3.}
\begin{equation}
  \phi=\frac{M}{r}+\frac{M_i x^i}{r^3}+\frac{M_{ij} x^i x^j}{r^5}+\ldots,
  \label{newtonian}
\end{equation}
where $r=(x^i x^i)^{\frac12}$, and the {\it multipole tensors}
$\{M_{ij\ldots k}\}$ are constant tensors defined in terms of
derivatives of the potential, evaluated at the origin ${\cal O}$. 
Upon shifting to a different point, the multipole moments change in a
way that involves only lower-order multipoles. The extension of these
ideas to General Relativity has been discussed in
refs.~\cite{Geroch:1970cc,Geroch:1970cd,Hansen:1974zz}, for general
asymptotically flat spacetimes. We will not need the full complication
of the latter, given that we will be concerned with solutions of the
massless free field equation of eq.~(\ref{masslessfreefield}) in
Minkowski space. Given a constant unit timelike vector $t^a$, one may
then consider the 3-space orthogonal to this, with induced metric
\begin{equation}
  h_{ab}=\eta_{ab}-t_a t_b.
  \label{hdef}
\end{equation}
Reference~\cite{Geroch:1970cc} then showed that an appropriate
generalisation of the multipole tensors appearing in
eq.~(\ref{newtonian}) is provided by symmetric, trace-free tensor
fields $Q^{a_1\ldots a_n}(x)$ satisfying
\begin{equation}
  t_{a_1} Q^{a_1\ldots a_n}=0,\quad
  \nabla^m Q^{a_1\ldots a_n}=\frac{n(2n-1)}{3}
  h^{m(a_1}Q^{a_2\ldots a_n)}
  -\frac{n(n-1)}{3}Q^{m(a_3\ldots a_n}h^{a_1a_2)},
  \label{multipoles}
\end{equation}
where the $n^{\rm th}$ such quantity is referred to as the {\it
  $2^n$-multipole tensor}, and the second condition requires that the
derivatives of multipole tensors depend only upon lower
multipoles. This is the analogue of the ``shifting the origin''
property mentioned for Newtonian multipoles above, and ensures that
the set of tensors $\{Q^{a_1\ldots a_n}\}$ corresponds to the same
solution of the field equation.  \\

In the spinorial formalism, each spacetime index in
eq.~(\ref{multipoles}) will become a pair of spinor
indices. Contracting with the timelike vector appearing there, one may
define the symmetric spinor field
\begin{equation}
  \omega^{A'_1\ldots A'_{2n}}=(6i)^n Q^{A'_1\ldots A'_n B_1\ldots B_n}
  t_{B_1}^{A'_{n+1}}\ldots t_{B_n}^{A'_{2n}},
\label{omegadef}
\end{equation}
which turns out to solve a higher-rank generalisation of the twistor
equation of eq.~(\ref{twistoreq}):
\begin{equation}
\nabla_L^{(L'}\omega^{A'_1\ldots A'_{2n})}=0.
\label{twistoreq2}
\end{equation}
Thus, we can associate the spinors of eq.~(\ref{omegadef}) with
multi-index {\it multipole twistors} $\{Q_{\alpha_1\ldots
  \alpha_{2n}}\}$. To see how this works in practice, consider a given
physical spin-$n$ field $\Psi_{A_1 A_2\ldots A_{2n}}$. Then one may
define higher-spin fields iteratively by taking derivatives and
contracting with the timelike vector appearing in
eq.~(\ref{omegadef}):
\begin{equation}
\Psi^{(n)}_{A_1\ldots A_{2n}}=t_{A'A_1}\nabla^{A'}_{A_2}[
  \Psi^{(n-1)}_{A_3\ldots A_{2n}}].
\label{Psindef}
\end{equation}
These constitute a spinorial analogue of the multiple derivatives
appearing in the Newtonian formalism of eq.~(\ref{Psindef}), whereby
higher multipole moments contain more derivatives of the original
potential. For a spin-1 field, one may write an explicit twistor space
integral for the total conserved charge producing the
field~\cite{Isham:1975dd}:
\begin{equation}
  Q=-\frac{i}{4\pi^2}\oint f(W_\alpha)d^4 W,\quad
  d^4 W=\frac{1}{4!}\epsilon^{\alpha\beta\gamma\delta}dW_\alpha\,dW_\beta
  \,dW_\gamma\,dW_\delta,
  \label{Qint}
\end{equation}
where an appropriate contour must be chosen, and where $f(W_\alpha)$
is the twistor function corresponding to the spacetime field. Given a
higher-spin field as in eq.~(\ref{Psindef}), we can form multiple
spin-1 fields by contracting with solutions of the twistor
equation\footnote{That the fields of eq.~(\ref{nPhiAB}) indeed
satisfy the massless free field equation of
eq.~(\ref{masslessfreefield}) follows from eq.~(\ref{twistoreq2}).}
$\{\alpha^{A_1\ldots A_{2n}}\}$:
\begin{equation}
\Phi^{(n)}_{AB}=-i^n \alpha^{A_1\ldots A_{2n}}[\Psi^{(n+1)}_{ABA_1
\ldots A_{2n}}].
\label{nPhiAB}
\end{equation}
Each of these fields will have a conserved charge according to
eq.~(\ref{Qint}), and we may collect together all such charges in the
twistor-covariant form
\begin{equation}
q(A^{\alpha_1\ldots \alpha_{2n}})=\frac{i^{n+1}}{4\pi^2}
\oint W_{\alpha_1}\ldots W_{\alpha_{2n}} A^{\alpha_1\ldots \alpha_{2n}}
f_{n+1}(W_\alpha)d^4 W,
\label{qmulti}
\end{equation}
for symmetric twistors $\{A^{\alpha_1\ldots\alpha_{2n}}\}$, where
$f_{n+1}(W_\alpha)$ is the twistor function corresponding to the
spacetime higher-spin field ${_{(n+1)}}\Psi_{ABA_1\ldots A_{2n}}$, and
multipole index $n$. Equation~(\ref{qmulti}) defines a set of
quantities dual to the $\{A^{\alpha_1\ldots\alpha_{2n}}\}$:
\begin{equation}
Q_{\alpha_1\ldots \alpha_{2n}}=\frac{i^{n+1}}{4\pi^2}
\oint W_{\alpha_1}\ldots W_{\alpha_{2n}}
f_{n+1}(W_\alpha)d^4 W,
\label{Qmulti}
\end{equation}
which are the multipole twistors we have been seeking. Note that the
iterative structure of the higher-spin fields in eq.~(\ref{Psindef})
implies that the twistor functions $\{f_{n+1}\}$ in eq.~(\ref{Qmulti})
can also be constructed iteratively, and there are various ways that
this can be written. A fully invariant condition is~\cite{Curtis}
\begin{equation}
f_n=i(R_\alpha W_\beta I^{\alpha\beta})^{-1} R_\gamma P^\gamma_\delta
\frac{\partial f_{n-1}}{\partial W_\delta},
\label{fniterate}
\end{equation}
where we have introduced the so-called {\it infinity twistors} for
Minkowski spacetime:
\begin{equation}
  I_{\alpha\beta}=\left(\begin{array}{cc}
  0 & 0 \\ 0 & \epsilon^{A'B'}\end{array}\right),\quad
  I_{\alpha\beta}=\left(\begin{array}{cc}
   \epsilon^{AB} & 0 \\ 0 &0\end{array}\right),
  \label{Idef}
\end{equation}
$R_\alpha$ is an arbitrary twistor, and we have introduced the
projector~\cite{Curtis}
\begin{equation}
  \lambda P^\alpha_\beta=I^{\alpha\gamma} Q_{\gamma\beta},
  \label{Prel}
\end{equation}
where $\lambda$ is the relevant mass or charge parameter for a given
theory. One thus has $\lambda=m$ in gravity, where $m$ is the total
mass of the system. In gauge or biadjoint theory, it will be the total
charge of the system that is creating the field, which we will denote
by $q$ and $y$ respectively.

\subsection{The double copy of the multipole expansion}
\label{sec:doublecopy}

\begin{figure}
\begin{center}
\scalebox{0.8}{\includegraphics{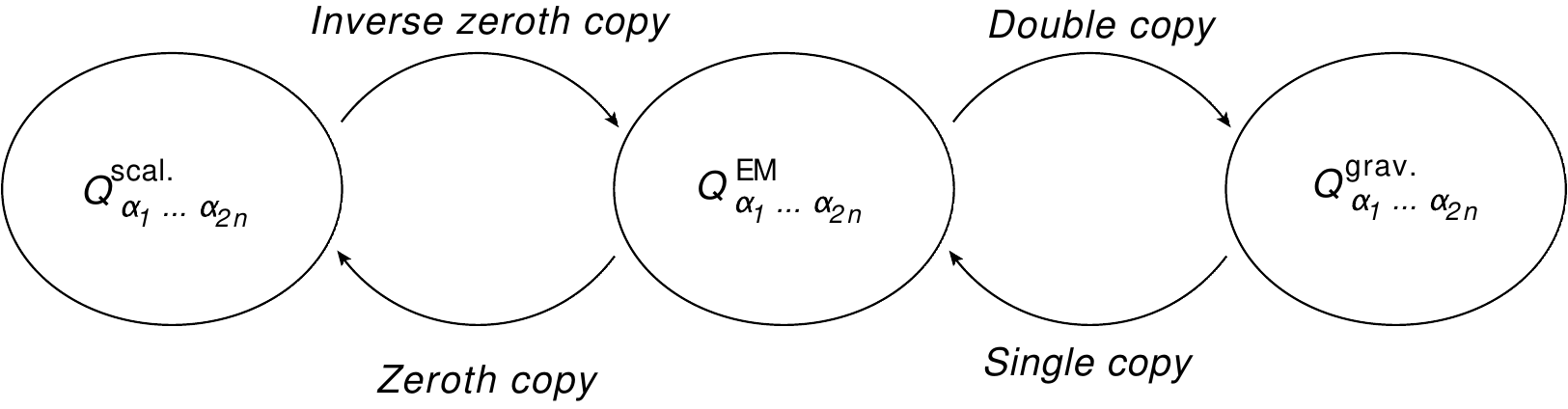}}
\caption{Double copy structure of the multipole twistors appearing in
  a single column in table~\ref{tab:multipoles2}.}
\label{fig:multipole_copies}
\end{center}
\end{figure}

The multipole twistors introduced above allow us to address the double
copy of the multipole expansion in a particularly compact and elegant
way. Consider twistor functions corresponding to a biadjoint scalar,
electromagnetic and gravity solution respectively, which we label by
$f_X(W_\alpha)$, $X\in\{{\rm scal.},{\rm EM},{\rm grav.}\}$. From each
of these, one may define a set of higher-spin twistor functions
according to the iterative procedure of eq.~(\ref{fniterate}), denoted
here by $f_{X}^{(n)}(W_\alpha)$. By eq.~(\ref{Qmulti}), this
immediately leads to a set of multipole twistors for each original
spacetime field. This construction is shown in
table~\ref{tab:multipoles}, where each column contains twistor
functions of the same homogeneity, leading to the same spin fields in
position space. We show the multipole twistors that arise from these
in table~\ref{tab:multipoles2}. Note that a gravitational monopole
contribution is not directly obtained from the corresponding twistor
functions. As the total mass, however, it is obtainable from the
angular momentum twistor $Q^{\rm grav.}_{\alpha_1\alpha_2}$. For the
$n=1$ case, one finds integrals expressing the total charge generated
by the biadjoint or EM field. From $n=2$ upwards, there are multipole
twistors in all three theories. For a given set of functions $\{f_X\}$
related by the twistor-space double copy, we can then associate each
column of table~\ref{tab:multipoles2} with a classical double copy
triple, as shown in figure~\ref{fig:multipole_copies}.\\
\begin{table}
\begin{center}
\begin{tabular}{c|ccccc}
Theory & \multicolumn{5}{c}{Multipole index $n\rightarrow $}\\
\hline
Biadjoint scalar & $f_{\rm scal.}$ & $f^{(1)}_{\rm scal.}$ &
 $f^{(2)}_{\rm scal.}$ &  $f^{(3)}_{\rm scal.}$ &  $f^{(4)}_{\rm scal.}$ \\
Gauge &  & $f_{\rm EM}$ &
 $f^{(2)}_{\rm EM}$ &  $f^{(3)}_{\rm EM}$ &  $f^{(4)}_{\rm EM}$ \\
Gravity &  &  &
 $f_{\rm grav.}$ &  $f^{(3)}_{\rm grav.}$ &  $f^{(4)}_{\rm grav.}$ 
\end{tabular}
\caption{Twistor functions occuring in the multipole expansion for
  three different theories. Here $f_X$ is a twistor function
  corresponding to physical spacetime field in theory $X$, and
  $f_X^{(n)}$ is a higher-spin field generated from this by the
  iterative procedure of eq.~(\ref{fniterate}).}
\label{tab:multipoles}
\end{center}
\end{table} 
\begin{table}
\begin{center}
\begin{tabular}{c|cccc}
Theory & \multicolumn{4}{c}{Multipole index $n$}\\
& 0 & 1 & 2 & 3\\
\hline
Biadjoint scalar & $Q^{\rm scal.}$ & $Q^{\rm scal.}_{\alpha_1\alpha_2}$ &
  $Q^{\rm scal.}_{\alpha_1\alpha_2\alpha_3\alpha_4}$ 
&  $Q^{\rm scal.}_{\alpha_1\ldots \alpha_6}$\\
Gauge & $Q^{\rm EM}$ & $Q^{\rm EM}_{\alpha_1\alpha_2}$ &
  $Q^{\rm EM}_{\alpha_1\alpha_2\alpha_3\alpha_4}$ 
&  $Q^{\rm EM}_{\alpha_1\ldots \alpha_6}$\\
Gravity &  & $Q^{\rm grav.}_{\alpha_1\alpha_2}$ &
  $Q^{\rm grav.}_{\alpha_1\alpha_2\alpha_3\alpha_4}$ 
&  $Q^{\rm grav.}_{\alpha_1\ldots \alpha_6}$
\end{tabular}
\caption{Multipole twistors arising from the twistor functions of
  table~\ref{tab:multipoles}.}
\label{tab:multipoles2}
\end{center}
\end{table} 

The physical interpretation of the identifications in
figure~\ref{fig:multipole_copies} is straightforward. Consider, for
example, the 2-multipole tensors $Q^{X}_{\alpha\beta}$. This
represents the angular momentum in gravity~\cite{Penrose:1972ia},
whereas in electromagnetism it is the charge dipole tensor, as
expected given that the single copy turns mass into charge. Likewise,
for the higher multipoles, the single copy replaces the relevant
spatiotemporal distribution of mass / momentum with that of charge,
with a further replacement to ``biadjoint charge'' in the zeroth
copy. \\

It is one thing to formally identify the multipole twistors in
different theories, as we have done in
figure~\ref{fig:multipole_copies}. It is quite another thing to say
that the multipole twistors in the different theories {\it are the
  same}, up to simple mass and charge replacements. Remarkably, this
strong statement indeed turns out to be true for fields related by the
original type D Weyl double copy of ref.~\cite{Luna:2018dpt}, as we
discuss in the following section.

\subsection{Multipole moments of type D solutions}
\label{sec:KerrMultipoles}

As discussed in ref.~\cite{Haslehurst} and reviewed in
refs.~\cite{White:2020sfn,Chacon:2021wbr}, all vacuum type D solutions
arise from twistor functions of the form
\begin{equation}
  f_{\rm grav.}=(A^{\alpha\beta}W_\alpha W_\beta)^{-3},
\label{fKerr1}
\end{equation}
where $A^{\alpha\beta}$ is a constant twistor that can be taken to be
symmetric. We see that eq.~(\ref{fKerr1}) has homogeneity $-6$ under
rescalings of $W_\alpha$, as required for a gravity
solution. Furthermore, it has two poles in twistor space, which give
rise, after performing the Penrose transform of eq.~(\ref{Penrose2})
to position space, to the two two-fold degenerate principal spinors of
the Weyl spinor that characterise it as being of type D. It turns out
that the twistor $A^{\alpha\beta}$ can be straightforwardly related to
the 2-multipole twistor for this field. Substituting
eq.~(\ref{fKerr1}) into eq.~(\ref{Qmulti}) for $n=1$, one may carry
out the integral using a special case of
\begin{equation}
\oint W_{\alpha_1}\ldots W_{\alpha_{2n}}
(W_\alpha W_\beta A^{\alpha\beta}))^{-(n+2)}d^4 W
=\frac{\pi^3 i}{\Delta}\frac{(2n)!}{2^{2n-1}(n+1)!n!}
B_{(\alpha_1\alpha_2}\ldots B_{\alpha_{2n-1}\alpha_{2n})},
\label{twistorint}
\end{equation}
where $B_{\alpha\beta}$ is the inverse of $A^{\alpha\beta}$, and
$\Delta$ the determinant of the latter. One finds
\begin{equation}
  Q_{\alpha\beta}=\frac{\pi}{8i\Delta}B_{\alpha\beta},\quad
  Q^{\alpha\beta}=\frac{8i\Delta}{\pi}A^{\alpha\beta}.
  \label{QKerr}
\end{equation}
Given the general type D gravity twistor function of
eq.~(\ref{fKerr1}), one may also identify the single and zeroth
copies, giving rise to a gauge and biadjoint scalar field in spacetime
respectively. As explained in
refs.~\cite{White:2020sfn,Chacon:2021wbr}, these are
\begin{equation}
  f_{\rm scal.}={\cal N}_0(A^{\alpha\beta}W_\alpha W_\beta)^{-1},\quad
  f_{\rm EM}={\cal N}_1(A^{\alpha\beta}W_\alpha W_\beta)^{-2}.\quad
 \label{fKerr2}
\end{equation}
We have here included arbitrary constant normalisation factors in the
scalar and electromagnetic functions, which are in any case not fixed
in the Weyl double copy of ref.~\cite{Luna:2018dpt}. Physically, one
may absorb such constants by redefining the total amount of charge in
a particular solution, but we will fix them shortly. Let us now
construct and compare the multipole twistors from these solutions. For
each field, we may construct higher-spin twistor functions using the
procedure of eq.~(\ref{fniterate}). Starting with the gravity function
from eq.~(\ref{fKerr1}), one finds
\begin{align}
  f^{(3)}_{\rm grav.}
&=-\frac{3i}{\alpha} (R_\alpha W_\beta I^{\alpha\beta})^{-1}
  (A^{\rho\lambda} W_\rho W_\lambda)^{-4}
 R_\gamma I^{\gamma\tau}Q_{\tau \delta}
  A^{\delta\sigma} W_\sigma,
  \label{f1calc1}
\end{align}
where we have used eq.~(\ref{Prel}). We may now use eq.~(\ref{QKerr}),
which yields
\begin{align}
  f^{(3)}_{\rm grav.}&=3\left(-\frac{\pi}{4\Delta m}\right) 
  (A^{\rho\lambda} W_\rho W_\lambda)^{-4},
  \label{f1calc2}
\end{align}
such that iterating this procedure leads to the formula
\begin{equation}
f^{(n)}_{\rm grav.}=\left(-\frac{\pi}{4\Delta m}\right)^{n-2}
\frac{n!}{2}(W_\alpha W_\beta A^{\alpha\beta})^{-(n+1)},
\label{fngrav}
\end{equation}
as quoted in ref.~\cite{Curtis}. Note that placing $n=2$ in this
formula reproduces the original gravity twistor function $f_{\rm
  grav.}(W_\alpha)$ itself. We may find the multipole twistors of
eq.~(\ref{Qmulti}) using eq.~(\ref{twistorint}, \ref{QKerr}), yielding
\begin{equation}
Q^{\rm
  grav.}_{\alpha_1\ldots\alpha_{2n}}=\frac{1}{2}\frac{1}{(2m)^{n-1}}
\frac{(2n)!}{n!}Q^{\rm grav.}_{(\alpha_1\alpha_2}\ldots Q^{\rm
  grav.}_{\alpha_{2n-1}\alpha_{2n})},
\label{Qcalc1}
\end{equation}
In principle, one may convert these multipole twistors back into
multipole tensors. For the Kerr solution, a set of scalar {\it
  multipole moments} has been defined in the
literature~\cite{Hansen:1974zz}. Let $\tilde{z}^a$ be a vector aligned
with the axis of rotation of the black hole, and $\Lambda$ be the
point at infinity after conformal compactification of the
spacetime. Then the multipole moments are given by
\begin{equation}
Q_n=\frac{1}{n!}Q_{a_1\ldots a_n}\tilde{z}^{a_1}\ldots z^{a_n}\Big|_\Lambda,
\label{Qndef}
\end{equation}
where the notation on the right-hand side denotes that this be
evaluated at $\Lambda$ itself. As stated in ref.~\cite{Curtis}, the
multipole twistors of eq.~(\ref{Qcalc1}) do indeed reproduce the known
multipole moments of the Kerr solution, first found in
ref.~\cite{Hansen:1974zz}.\\

We may carry out the above procedure for the biadjoint scalar and
gauge theory twistor functions of eq.~(\ref{fKerr2}), and the resulting
higher spin twistor functions are given by
\begin{align}
f^{(n)}_{\rm scal.}&={\cal N}_0 
\left(-\frac{\pi}{4\Delta y}\right)^{n}
\frac{n!}{2}(W_\alpha W_\beta A^{\alpha\beta})^{-(n+1)}\notag\\
f^{(n)}_{\rm EM}&={\cal N}_1 
\left(-\frac{\pi}{4\Delta q}\right)^{n-1}
\frac{n!}{2}(W_\alpha W_\beta A^{\alpha\beta})^{-(n+1)},
\label{fn01}
\end{align}
where we have replaced the mass $m$ in the gravity solution with the
charge $q$ in gauge theory, and biadjoint charge $y$ in the scalar
theory. These functions reproduce the original fields for $n=0$ and
$n=1$ respectively. We may choose to fix the arbitrary normalisation
constants ${\cal N}_i$ by requiring that the 2-multipole (dipole)
tensor in each theory is simply related by replacing 
\begin{equation}
m\rightarrow q\rightarrow y.
\label{mreplace}
\end{equation}
in going from gravity to gauge theory, to biadjoint theory. This
determines
\begin{equation}
{\cal N}_0=\left(-\frac{\pi}{4\Delta y}\right)^{-2},\quad
{\cal N}_1=\left(-\frac{\pi}{4\Delta q}\right)^{-1},
\label{N0N1}
\end{equation}
after which comparison of eq.~(\ref{fn01}) with eq.~(\ref{fngrav})
shows that all higher-spin twistor functions agree across all three
theories, so that one may simply replace the multipole twistors of
eq.~(\ref{Qcalc1}) with the gauge and biadjoint scalar counterparts
\begin{align}
Q^{\rm scal.}_{\alpha_1\ldots\alpha_{2n}}&=\frac{1}{2}\frac{1}{(2y)^{n-1}}
\frac{(2n)!}{n!}Q^{\rm scal.}_{(\alpha_1\alpha_2}\ldots 
Q^{\rm scal.}_{\alpha_{2n-1}\alpha_{2n})}\notag\\
Q^{\rm EM}_{\alpha_1\ldots\alpha_{2n}}&=\frac{1}{2}\frac{1}{(2q)^{n-1}}
\frac{(2n)!}{n!}Q^{\rm EM}_{(\alpha_1\alpha_2}\ldots 
Q^{\rm EM}_{\alpha_{2n-1}\alpha_{2n})},
\label{Qcalc2}
\end{align}
As a direct consequence, the multipole moments of the gauge and
biadjoint scalar fields corresponding to a given gravity field from
eq.~(\ref{fKerr1}) precisely match, after making the necessary
mass-to-charge replacements. Our twistor analysis has applied for an
arbitrary quadratic form in eq.~(\ref{fKerr1}) which, as explained in
refs.~\cite{White:2020sfn,Chacon:2021wbr}, is a general statement for
any (vacuum type D) spacetime entering the original Weyl double copy
of ref.~\cite{Luna:2018dpt}. In particular, this must apply to the
Kerr solution, and there is a novel cross-check one may perform. The
gauge theory counterpart of this solution is the $\sqrt{\rm Kerr}$
solution discussed above, and its electromagnetic monopole moments
have not previously been calculated directly. However, one may instead
consider a charged Kerr black hole, otherwise known as a Kerr-Newman
black hole~\cite{Newman:1965tw,Newman:1965my}. This is a solution of
the Einstein-Maxwell equations, and as such consists of a metric plus
a gauge field. The $\sqrt{\rm Kerr}$ solution can be obtained by
setting the mass of the solution to zero, leaving a gauge field living
in Minkowski space, and which is known to correspond to the single
copy of the gravity solution. The combined gravitational and
electromagnetic multipole moments of the Kerr-Newman solution have
been calculated in ref.~\cite{Sotiriou:2004ud}. The gravity moments
agree with the pure Kerr solution, and the electromagnetic ones are
simply obtained by the mass-to-charge replacement of
eq.~(\ref{mreplace}). This indeed verifies our results. Further
support comes from recent studies showing that the multipole expansion
for $\sqrt{\rm Kerr}$ and Kerr can be obtained from scattering
amplitudes that manifestly double-copy~\cite{Guevara:2020xjx}.

\section{Discussion}
\label{sec:discuss}

In this paper, we have considered whether the multipole expansions of
fields in biadjoint scalar, gauge and gravity theory can be related by
the classical double copy. By combining a twistor formulation of the
multipole expansion~\cite{Curtis} with a recently developed twistor
language for the classical double
copy~\cite{White:2020sfn,Chacon:2021wbr}, we have shown that the
multipole moments for arbitrary type D vacuum solutions indeed match
up in different theories, subject to appropriate mass / charge
replacements.\\

Our results provide a nice illustration of the efficiency of the
twistor double copy, but are of interest in their own right. It is
often the case that a single copy of a given gravity solution can be
found, but not easily interpreted. A canonical case of this is the
single copy of the Kerr black hole, first formally identified in
ref.~\cite{Monteiro:2014cda}, and denoted as $\sqrt{\rm Kerr}$ in
subsequent literature (see
e.g. refs.~\cite{Arkani-Hamed:2019ymq,Emond:2020lwi,Guevara:2020xjx}). It
is known that this solution occurs by replacing the source for the
Kerr black hole (a rotating disk of mass) with a similar gauge theory
source (a rotating disk of charge). However, the nature of the sources
is subtly different in the two theories~\cite{Monteiro:2014cda}, such
that it is not clear what impact this has on the fields
themselves. Multipole moments, however, allow us to fully characterise
the structure of fields in a gauge-invariant way. Thus, the fact that
the multipole moments for the Kerr and $\sqrt{\rm Kerr}$ solutions are
essentially identical tells us a great deal of information about how
to physically interpret the single copy, by recycling our intuition
gathered from the Kerr black hole. Furthermore, the fact that our
results apply for any type D vacuum solution makes this a rather
powerful statement, that may well help in interpreting and extending
the double copy in future.\\

Another nice aspect of our results is that the multipole expansion in
biadjoint theory also matches that in the gauge and gravity theories,
for the wide class of solutions we have considered. This adds a
powerful weight to the observations made in
refs.~\cite{White:2020sfn,Chacon:2021wbr}, namely that the twistor
double copy allows us to understand the {\it inverse zeroth copy} from
biadjoint scalar theory to gauge theory. That is, we have seen
directly that the multipoles of vacuum type D gravity solutions and
their single copies are essentially inherited directly from a much
simpler scalar theory! It is interesting to ponder what other physical
quantities can be phrased in such an appealing manner.

\section*{Acknowledgments}

We are very grateful to Donal O'Connell and Justin Vines for
discussions. This work has been supported by the UK Science and
Technology Facilities Council (STFC) Consolidated Grant ST/P000754/1
``String theory, gauge theory and duality'', and by the European Union
Horizon 2020 research and innovation programme under the Marie
Sk\l{}odowska-Curie grant agreement No. 764850 ``SAGEX''. EC is
supported by the National Council of Science and Technology (Conacyt).
AL is supported by the U.S. Department of Energy (DOE) under award
number DE-SC0009937, and by the Mani L. Bhaumik Institute for
Theoretical Physics.

\bibliography{refs}
\end{document}